\documentclass[journal]{IEEEtran}
\usepackage{amsfonts}
\usepackage{amsmath}
\usepackage{amssymb}
\usepackage{graphicx}

\begin{document}

\title{Dynamics of a bistable VCSEL subject to optical feedback from a vibrating rough surface}
\author{V.~N.~Chizhevsky
\thanks{Manuscript received. This work was supported by Belarus State
Programme for Scientific Research ``Photonics and opto- microeletronics\textquotedblright{}
(Task 2.1.01).}
\thanks{V.~N.~Chizhevsky is with B.~I. Stepanov Institute of Physics, NAS, Belarus (e-mail: vnc@dragon.bas-net.by).}
}

\maketitle

\maketitle
\begin{abstract}

The paper presents an experimental study of the temporal behaviour of a bistable vertical cavity surface emitting laser (VCSEL) under the effect of optical feedback coming from vibrating diffusely-reflecting surface. We demonstrate that a VCSEL operating in the regime of polarization switchings caused by self-mixing effects can greatly enhance a responsivity of the detection of microvibrations. For small amplitudes of microvibrations (less than $\lambda$/2, where $\lambda$ is the operating wavelength of the laser), which cause only harmonic (or close to harmonic) oscillations in the laser output outside of the bistability  region, the use of polarization switching can increase the responsivity up to 45 times. For the amplitudes larger then $\lambda$/2 the response of a  bistable VCSEL well reproduces a fine temporal structure of the self-mixing signal. A procedure of the data processing in the switching regime for the correct determination of the direction of the displacement and for the reconstruction of the waveform of surface vibrations with a basic resolution of $\lambda$/2 is also shown.

\end{abstract}


%\section{Introduction}
Vertical cavity surface emitting lasers subject to impact of optical feedback were a subject of wide experimental and theoretical interests in the context of studying different aspects of nonlinear behavior (see, for instance, \cite{Panajotov2013} and references therein). One of these effects is an appearance of fluctuations of the laser intensity caused by the reinjection of the laser radiation from the reflecting  moving surface. This self-mixing (SM) effect lies in the basis for laser interferometric sensors which are known to be effective for contactless measurements of displacements, velocities, vibrations, the direction of displacement of objects in different fields of science and technics (\cite{Donati12,Taimre15,Donati2018} and references therein). In this method a reinjected radiation induces fluctuations in the output laser emission which contains an information about object under study. Such an approach makes possible to develop, based on semiconductor lasers, measuring sensors that have a high sensitivity to the reflected signal, a low cost and a compactness. Especially this concerns VCSELs which possess a number of advantages such as a quality of the beam, an operation in a single transverse and longitudinal mode, a very low laser threshold and a possibility to fabricate an array of the laser diodes.

In this context Porta \textit{et al}. \cite{Porta02} experimentally demonstrated that the use of a bistable VCSEL  where the polarization switchings are induced by the laser radiation scattered back into the laser cavity from a rotating diffusely-reflecting object can strongly enhanced the responsivity of the laser Doppler velocimetry (by 10 dB). These experimental findings were also theoretically analyzed in the framework of the Lang$-$Kobayashi approach \cite{Albert04}.

In this paper we extend the use of polarization switchings in a bistable VCSEL for increasing sensitivity of the detection of microvibrations of a diffusely-reflective surface.
Here we show that for  microvibrations of small amplitudes (less than $\lambda$/2)
the use of polarization switching allows one to increase the response amplitude at fundamental frequency of microvibrations from 10 up to 45 times depending on theirs amplitudes. For the large enough amplitudes (larger then $\lambda$/2)  the response of a  bistable VCSEL well reproduces a fine temporal structure of the SM signal. This fact is demonstrated with help of the coefficient of cross-correlation between the spectrum of oscillations outside of bistable region and the spectrum of the laser response in the switching regime in the bistability domain.
In the case of symmetrical configuration of the bistable potential associated with polarization bistability in a VCSEL the coefficient of cross-correlation can achieve the value of 0.96.
We also study the effect of asymmetry of a bistable potential of the laser on the the response of the laser. Finally we discuss a procedure of the data processing in the switching regime in order to find correctly the direction of the displacement and to reconstruct a waveform of surface vibrations with a basic resolution of $\lambda$/2. It should be noted that a considerable body of works were devoted to developments and investigations of SM vibrometers based on optical feedback \cite{Roos96,Guliani03,Scalise04,Donati06,Zabit10,Zhao13,Tao15}, which are very important sensors for many applications. Therefore, the results presented below can be of interest for enhancement of the responsivity of laser-based vibrometers.

\begin{figure}[tbp]
\begin{center}
\includegraphics[clip,width=6cm,height=3.5cm]{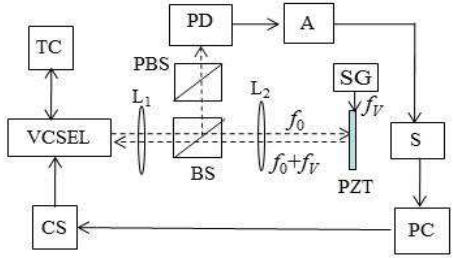}
\end{center}
\caption{Experimental setup: CS, current source; TC, temperature
controller; SG, signal generator; $L_{1}$ and $L_{2}$, lenses: HWP, half wave plate; BS, beam splitter: PBS, Glan prism; PD, photodiode; A, amplifier; S, USB digital oscilloscope; PZT, piezo-electric transducer; PC, computer.}
\label{fig1}
\end{figure}

Investigations were performed on the experimental setup schematically shown in Fig. \ref{fig1}.
\begin{figure}[tbp]
\begin{center}
\includegraphics[clip,width=6.2cm,height=10cm]{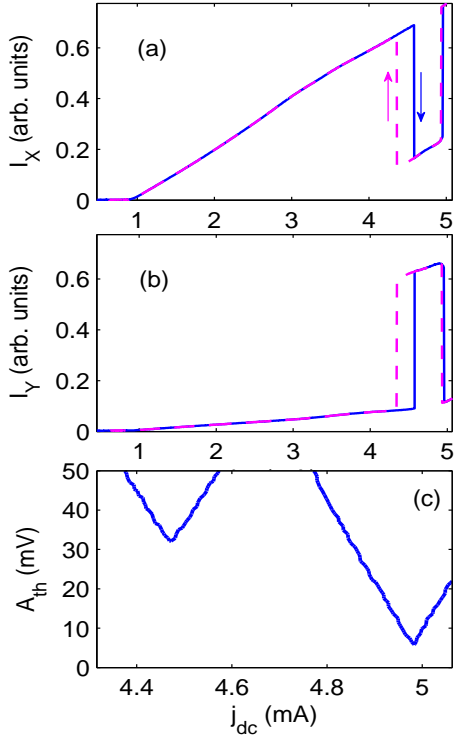}
\end{center}
\caption{Polarization resolved laser intensity versus injection current $j_{dc}$ (LI-curve) of a VCSEL shown for two orthogonal polarizations $I_{X}$ (a) and $I_{Y}$ (b), respectively. The blue line in both figures corresponds to an increase of the injection current, whereas the magenta dashed line denotes the opposite case. (c) The dependence of the switching threshold between polarization states in bistability zones as a function of the injection current $j_{dc}$ for the current modulation at the frequency $f$=1kHz.}
\label{fig2}
\end{figure}
A commercial single-mode VCSEL (Finisar, HFE 4093-332) operating at 848nm wavelength was used. The threshold current was $j_{th}\approx$0.9 mA. The temperature of the laser diode was stabilized with the help of a temperature controller with an accuracy of $0.01^\circ$C. The collimated by lens $L_{1}$ with an antireflection coating laser radiation was split on the beam splitter (BS) into two parts.
One of them was focused by a lens $L_{2}$ on a diffuse-reflected printer white paper which was fixed on a piezoelectric transducer (PZT) as a vibrating surface. Focal lengths of both lenses were 8 mm and 20 mm, respectively. The second beam was directed through the polarizer (PBS) to the photodiode (PD) to record the temporal dynamics on the selected polarization. The signal from the photodiode was amplified and recorded by a digital USB oscilloscope (with a sampling frequency of 1$\div$10 MHz). The length of the external cavity from the laser to the target was $\approx$12 cm.
For the modulation of the length of the optical feedback, a sinusoidal voltage with a frequency of 1 kHz and different amplitudes $U$ was applied to the piezoelectric transducer from the signal generator.

First, for several samples of laser diodes the dependence of the laser intensity $I(t)$ on the selected polarization depending on the current $j_{dc}$ was experimentally investigated. The hysteretic behavior of the laser intensity for high values of $j_{dc}$ was found for one of the laser diodes at successive increase and decrease of $j_{dc}$.  Figures \ref{fig2} (a) and (b) show laser intensities on the selected orthogonal polarizations for this case. In particular, in the range of $j_{dc}$ from $\approx4.3$ mA to $\approx4.9$ mA two zones of bistability with width $\Delta j_{1}\simeq$ 0.2 mA and $\Delta j_{2}\simeq$ 0.03 mA, respectively, were found. In the absence of optical feedback for a fixed value of $j_{dc}$ no spontaneous switchings caused by internal laser noise is observed in the operation of the laser in a bistable mode in both zones of bistability.

The switching threshold to the effect of the current modulation is significantly lower in the second zone [Fig. \ref{fig2}(c)] by about 8 times for the symmetrical configuration of a bistable potential associated with polarization bistability. In the experiment it can be found as a minimal modulation amplitude of a switching threshold in the dependence on the injection current $j_{dc}$.  In this case switching thresholds are the same from both potential wells. Changing the pump current from the symmetrical configuration, we introduce an additional threshold from one of the wells which results in increase of the switching threshold for the periodic signal.
One can expect that the same effect will be observed to the impact of the modulation of optical feedback. The second zone of bistability, therefore, was used in the experimental study. In order to find switchings between polarization states induced by optical feedback from the rough surface the PZT with a sinusoidal applied voltage was moved along the optical axis until clear-cut polarization switchings appear in the laser responses.

\begin{figure*}[!htbp]
\begin{center}
\includegraphics[clip,width=15cm,height=10cm]{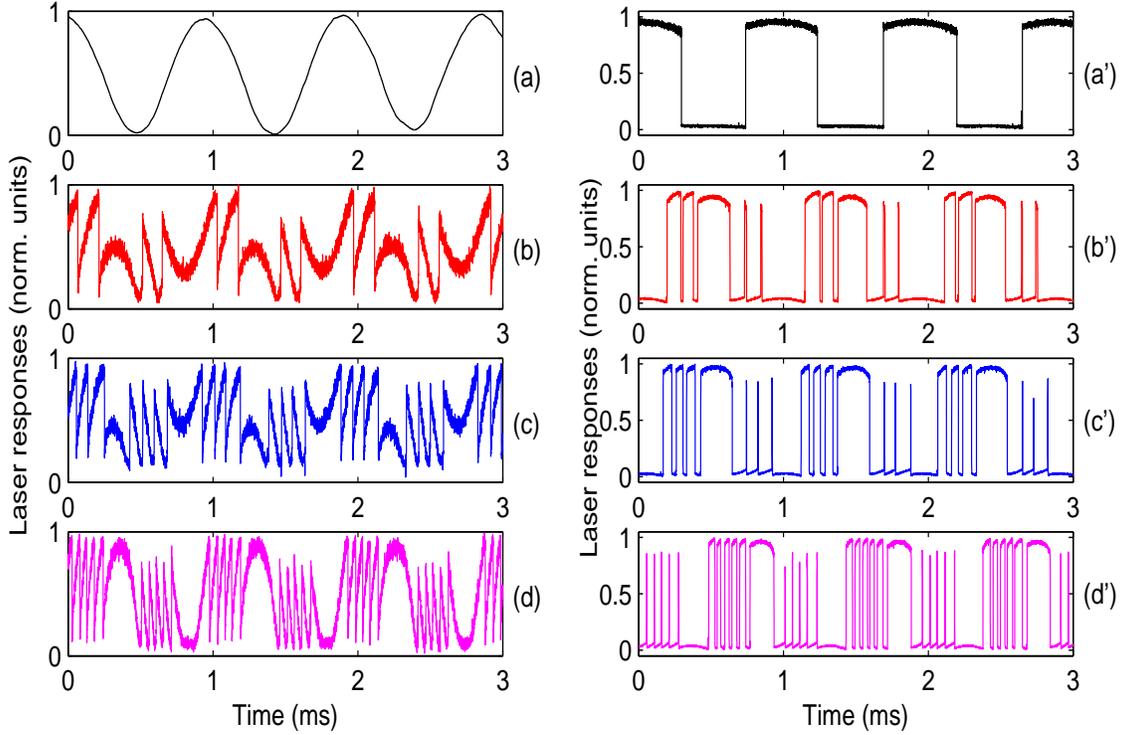}
\end{center}
\caption{Left column: Normalized SM waveforms obtained with a laser below the switching region shown for different values of the voltage on the PZT: 1 (a), 3 (b), 5 (c), 7 (d)  Volts. $j_{dc}$ = 4.873 mA. Right column: Corresponding temporal laser responses showing switchings between polarization states in the bistability region for the same applied voltages on the PZT. $j_{dc}$ = 4.975 mA. }
\label{fig3}
\end{figure*}

It is known that the reflected radiation returned back to the laser cavity from the vibrating object creates beats on the vibration frequency, thereby modulating the output power of the laser. The character of fluctuations of the output laser intensity depends both on the amplitude of the vibrations and on the strength of the optical feedback.
The left column in Fig. \ref{fig3} shows typical SM waveforms of the laser intensity $I(t)$ on the selected polarization for different values of the applied voltage $U$ on the PZT with a frequency $f_{V}$ = 1 kHz. These responses were obtained for the injection current outside of the second bistability range. At the value $U$ = 1 Volt, we observe oscillations of the laser intensity which are close to the harmonic ones [Fig. \ref{fig3}(a)]. An increase of the voltage $U$ on PZT results in the appearance of fringe-discontinuities (sawtooth-like fringes) in the laser intensity caused by a change in the length of the external cavity by multiple number of half-wavelength of the laser radiation [Fig. \ref{fig3}(b)-\ref{fig3}(d), left column].

In the bistability region SM signals are significantly changed as seen in the right column in Fig.\ref{fig3}. For $U$ = 1 V on the PZT the harmonic oscillations are transformed almost into squarewave signals due to switchings between two polarization states [Fig.\ref{fig3}(a')]. For higher values of $U$ on the PZT more complex fringe patterns induce additional switchings [Fig. \ref{fig3}(b')-\ref{fig3}(d'), right column]. But what is important, the number of switchings appeared in the laser response in each  one-half period of the modulation in the bistability region is equal to the number of fringe-discontinuities. This means that a fine structure of the SM signal can be kept in a bistable mode of operation of the laser.
\begin{figure}[tpb]
\begin{center}
\includegraphics[clip,width=6cm,height=8cm]{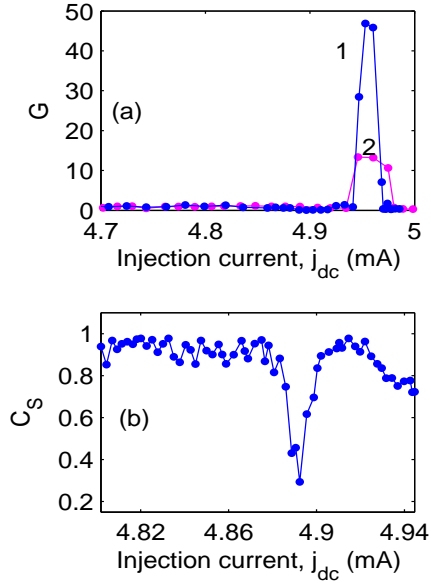}
\end{center}
\caption{(a) The gain factor $G$ versus the current magnitude $j_{dc}$ for the case of the symmetrical configuration of bistable potential (curves 1 and 2 correspond to the applied voltage on the PZT  0.25 V and 1 V, respectively). (b) Coefficient cross-correlation between $S_{0}(f)$ and the spectrum $S(f)$ for running value of $j_{dc}$ as a function of $j_{dc}$. Applied voltage on the PZT 8 Volts.}
\label{fig4}
\end{figure}

\begin{figure}[t]
\begin{center}
\includegraphics[clip,width=8.5cm,height=4.5cm]{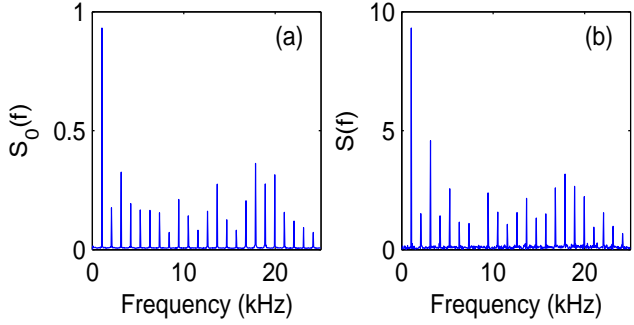}
\end{center}
\caption{(a) Amplitude spectrum $S_{0}(f)$ of the SM signal outside of the bistable region, (b) Amplitude spectrum $S_{j}(f)$ of the laser response at the optimal values of the $j_{dc}$ in the bistability region. The voltage on the PZT is 8 V.}
\label{fig5}
\end{figure}

\begin{figure}[!htp]
\begin{center}
\includegraphics[clip,width=7.5cm,height=6cm]{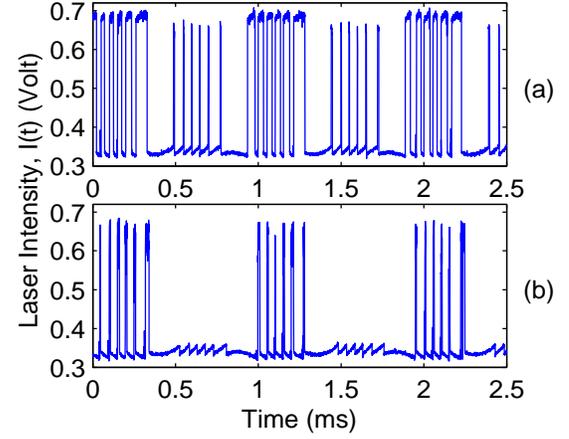}
\end{center}
\caption{ Temporal laser responses for the case of asymmetrical bistable potential.  (a) $j_{dc}$ = 4.967, (b) $j_{dc}$ = 4.985. The voltage on the PZT is 8 Volts.}
\label{fig6}
\end{figure}

For the harmonic or close to harmonic SM signal one can introduce the gain factor since in the this case a dominant frequency in the spectrum of the signal is $f_{V}$. Figure \ref{fig4} shows the gain factor $G$ for the signal at the vibration frequency $f_{V}$ depending on the value of the pump current $j_{dc}$ for two different values of sinusoidal voltage $U$ applied to the PZT. The gain factor $G$ is defined here as $G=R/R_{0}$, where $R$ is the response of the system at the frequency $f_{V}$ for the running value of $j_{dc}$, and $R_{0}$ is the average laser response outside of the bistability region for the range $j_{dc}=4.7\div4.8$ mA. It is seen that for small amplitude of vibrations outside of bistable region the gain factor $G$ does not depend on $j_{dc}$ whereas in the bistable regime of operation it can achieve a rather high value of about 45 for $U$=0.25 V [curve 1, Fig.\ref{fig4}(a)].
In the case of more complex signals [Fig. \ref{fig3}(d)], along with a frequency $f_{V}$ a lot of spectral components appear in the spectra of laser responses. To find out a similarity between them, therefore, a normalized cross-correlation coefficient ($C_{s}$) between the spectrum $S_{0}$ averaged in the range of $j_{dc}\in[4.8\div4.83]$ mA outside of bistable zone and the spectrum $S_{j}$ for each value of $j_{dc}$ was estimated:
\begin{equation}
C_{S}=\frac{\left\langle S_{0}S_{j}\right\rangle -\left\langle
S_{0}\right\rangle \left\langle S_{j}\right\rangle }{[(\left\langle
S^{2}_{0}\right\rangle -\left\langle S_{0}\right\rangle ^{2})(\left\langle
S^{2}_{j}\right\rangle -\left\langle S_{j}\right\rangle ^{2})]^{1/2}},
\label{eqn1}
\end{equation}
where $\left\langle ...\right\rangle$ denotes the average in the frequency domain. We used the frequency range from 0 up to 50$f_{V}$ which captures practically all frequency components of the SM signals in our study. The cross-correlation coefficient $C_{S}$ depending on the injection current $j_{dc}$, calculated with the use of expression (\ref{eqn1}), is shown on Fig. \ref{fig4}(b). One can note that outside of the bistable zone the values of $C_{S}$ are between 0.9  and 1 due to some fluctuations of the strength of the optical feedback and the length of the external cavity. In the bistability domain the coefficient $C_{S}$ can reach the value $C_{S}\approx0.96$ for some optimal value of $j_{dc}$. This fact testifies a high degree of similarity between the averaged spectrum $S_{0}$ (considering as an input optimal spectrum) and the output spectrum $S_{j}$. In fact, the optimal value of $j_{dc}$ for $G$ and $C_{S}$ are achieved for the injection current $j_{dc}$ corresponding to the symmetric configuration of the bistable potential. In this case the switching threshold is minimal. The spectra $S_{0}$ and $S_{j}$ at the optimal value of $j_{dc}$ are shown in the Figs.\ref{fig5} (a) and \ref{fig5}(b), respectively. Comparing both spectra one can note that practically all spectral components are well reproduced in the regime of polarization switchings, at the same time the amplitudes of spectral peaks increase approximately by a factor of 10.
As we noted yet, the changing the pump current from the symmetrical configuration, we introduce some level of asymmetry to a bistable potential which increases the switching threshold for the periodic signal. This results in a somewhat different picture as compared to results presented in Fig.\ref{fig3}. The experimental results for the temporal behavior of the laser presented in Figs. \ref{fig6}(a) and \ref{fig6}(b) are shown for two different values of $j_{dc}$ which correspond to two different level of asymmetry of the bistable potential.  For $j_{dc}$ = 4.967 mA [Fig. \ref{fig6}(a)] one can notice the appearance of bursts of pulses at the rate corresponding to the frequency $f_{V}$. At the same time, the width of pulses in each semi-period $T_{V}$ sequentially alternates. Further increase of the asymmetry of a bistable potential ($j_{dc}$ = 4.985 mA) results in a disappearance of pulses in one of semi-period of oscillations [Fig. \ref{fig6}(b)].
\begin{figure}[t]
\begin{center}
\includegraphics[clip,width=8.5cm,height=7cm]{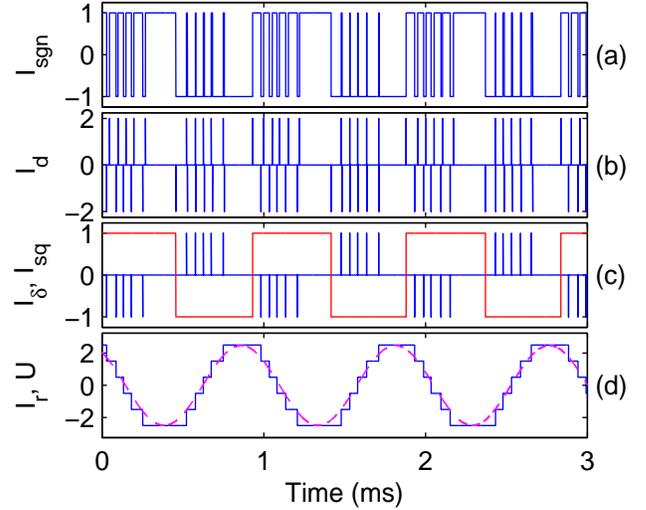}
\end{center}
\caption{ (a) The signal $I_{sgn}(t)$, (b) its derivative $I_{d}$, (c) the signal $I_{sq}(t)$ on the fundamental frequency $f_{V}$ and processed signal $I_{\delta}(t)$, (d) reconstructed waveform (stepwise signal) and the modulation signal on PZT (dashed line).}
\label{fig7}
\end{figure}

Finally, let us consider the possibility to reconstruct a waveform of vibrations from the SM signal in the switching regime. It is known that a SM interferometry allows one to measure the displacement of a target with a basic resolution of $\lambda$/2. The  procedure is based on counting of saw-tooth like fringes, which number is directly proportional to the displacement. The similar procedure can be developed in the switching regime, even though the SM signals are significantly different in both cases [compare Fig. \ref{fig3}(d) and Fig. \ref{fig3}(d')]. Let us consider a laser response in Fig. \ref{fig3}(d') (right column), which corresponds to a SM waveform $I(t)$ in the switching regime. First, we transform $I(t)$ into the signal $I_{sgn}(t)$ using the following expression: $I_{sgn}(t)=sgn[I(t)-\langle I(t)\rangle]$, where $sgn$ denotes a signum function (defined as 1 for positive $I(t)$, -1 for negative $I(t)$ and null at zero) and $\langle ..\rangle$ is an operation of averaging over a full length of time series. The result of this transformation is shown in Fig. \ref{fig7} (a).
Then we calculate derivative $I_{d}(t)=dI_{sgn}(t))/dt$ shown in Fig.\ref{fig7} (b). One can see that signal $I_{d}(t)$ represents an alternation of delta-peaks of the opposite sign due to a rectangular character of the signal $I_{sgn}(t)$. By selecting each first peak from the pair of delta-peaks of the opposite sign in each half-period of the fundamental frequency $f_{V}$ [solid thick line in Fig. \ref{fig7}(c)] we obtain a series of delta-peaks $I_{\delta}(t)$ of the same sign in each half-period of $f_{V}$ However, the sign of peaks is changed to the opposite value in each semiperiod of $f_{V}$. This change of sign can be associated with a change of the direction of a displacement as in a conventional SM interferometry. A distance between peaks corresponds to a change of the length of the external cavity by $\lambda/2$. To reconstruct the waveform of micro-vibrations we calculate a cumulative sum of $I_{\delta}(t)$. The stepwise curve shown in the Fig. \ref{fig7}(d) represents the reconstructed waveform with resolution $\lambda/2$. The dashed line is the signal from the functional generator applied to the PZT.

Thus, it was experimentally demonstrated that the amplitude of SM signal caused by optical feedback from vibrating diffusely-reflective surfaces can be significantly enhanced by the use of polarization switchings a bistable VCSEL by a factor of about 45 for the magnitude of microvibrations less than $\lambda/2$. In the more complicated case for the magnitude of microvibrations greater than $\lambda/2$, spectra of the laser response in the switching regime well reproduce all spectral peculiarities of self-mixing signals. For the case of the weak OF, when the strength of OF is not enough to induce polarizations switching, a further improvement can be achieved by the use of phenomenon of vibrational resonance \cite{Landa00,Chizhevsky03,Chizhevsky12}. Preliminary results of the use of vibrational resonance for the amplification of the SM signal was reported in \cite{Chizhevsky18}

\bibliographystyle{IEEEtran}

\begin{thebibliography}{10}
\providecommand{\url}[1]{#1}
\csname url@rmstyle\endcsname
\providecommand{\newblock}{\relax}
\providecommand{\bibinfo}[2]{#2}
\providecommand\BIBentrySTDinterwordspacing{\spaceskip=0pt\relax}
\providecommand\BIBentryALTinterwordstretchfactor{4}
\providecommand\BIBentryALTinterwordspacing{\spaceskip=\fontdimen2\font plus
\BIBentryALTinterwordstretchfactor\fontdimen3\font minus
  \fontdimen4\font\relax}
\providecommand\BIBforeignlanguage[2]{{%
\expandafter\ifx\csname l@#1\endcsname\relax
\typeout{** WARNING: IEEEtran.bst: No hyphenation pattern has been}%
\typeout{** loaded for the language `#1'. Using the pattern for}%
\typeout{** the default language instead.}%
\else
\language=\csname l@#1\endcsname
\fi
#2}}

\bibitem{Panajotov2013}
K.~Panajotov, M.~Sciamanna, M.~A. Arteaga, and H.~Thienpont, ``Optical feedback
  in vertical-cavity surface-emitting lasers,'' \emph{IEEE Journal of Selected
  Topics in Quantum Electronics}, vol.~19, no.~4, p. 1700312, 2013.

\bibitem{Donati12}
S.~Donati, ``Developing self-mixing interferometry for instrumentation and
  measurements,'' \emph{Laser Photonics Rev.}, vol.~6, no.~3, pp. 393--417,
  2012.

\bibitem{Taimre15}
T.~Taimre, M.~Nikoli\'{c}, K.~Bertling, Y.~L. Lim, T.~Bosch, and A.~D.
  Raki\'{c}, ``Laser feedback interferometry: a tutorial on the self-mixing
  effect for coherent sensing,'' \emph{Adv. Opt. Photon.}, vol.~7, no.~3, pp.
  570--631, 2015.

\bibitem{Donati2018}
S.~Donati and M.~Norgia, ``Overview of self-mixing interferometer applications
  to mechanical engineering,'' \emph{Optical Engineering}, vol.~57, pp.
  051\,506--051\,506--13, 2018.

\bibitem{Porta02}
P.~A. Porta, D.~P. Curtin, and J.~G. McInerney, ``Laser doppler velocimetry by
  optical self-mixing in vertical-cavity surface-emitting lasers,''
  \emph{Photonics Technology Letters, IEEE}, vol.~14, no.~12, pp. 1719--1721,
  2002.

\bibitem{Albert04}
J.~Albert, M.~C. Soriano, I.~Veretennicoff, K.~Panajotov, J.~Danckaert, P.~A.
  Porta, D.~P. Curtin, and J.~G. McInerney, ``Laser doppler velocimetry with
  polarization-bistable vcsels,'' \emph{Selected Topics in Quantum Electronics,
  IEEE Journal}, vol.~10, no.~5, pp. 1006--1012, 2004.

\bibitem{Roos96}
P.~A. Roos, M.~Stephens, and C.~E. Wieman, ``Laser vibrometer based on
  optical-feedback-induced frequencymodulation of a single-mode laser diode,''
  \emph{Appl. Opt.}, vol.~35, no.~34, pp. 6754--6761, 1996.

\bibitem{Guliani03}
G.~Giuliani, S.~Bozzi-Pietra, and S.~Donati, ``Self-mixing laser diode
  vibrometer,'' \emph{Measurement Science and Technology}, vol.~14, no.~1,
  p.~24, 2003.

\bibitem{Scalise04}
L.~Scalise, Y.~Yu, G.~Giuliani, G.~Plantier, and T.~Bosch, ``Self-mixing laser
  diode velocimetry: application to vibration and velocity measurement,''
  \emph{IEEE Trans. Instrumentation and Measurement}, vol.~53, pp. 223--232,
  2004.

\bibitem{Donati06}
S.~Donati, M.~Norgia, and G.~Giuliani, ``Self-mixing differential vibrometer
  based on electronic channel subtraction,'' \emph{Appl. Opt.}, vol.~45,
  no.~28, pp. 7264--7268, 2006.

\bibitem{Zabit10}
U.~Zabit, R.~Atashkhooei, T.~Bosch, S.~Royo, F.~Bony, and A.~D. Rakic,
  ``Adaptive self-mixing vibrometer based on a liquid lens,'' \emph{Opt.
  Lett.}, vol.~35, no.~8, pp. 1278--1280, Apr 2010.

\bibitem{Zhao13}
Y.~Zhao, L.~Lu, Z.~Du, B.~Yang, W.~Zhang, J.~Zhou, H.~Gui, and B.~Yu,
  ``Research on micro-vibration measurement by a laser diode self-mixing
  interferometer,'' \emph{Optik - International Journal for Light and Electron
  Optics}, vol. 124, no.~21, pp. 4707--4711, 2013.

\bibitem{Tao15}
Y.~Tao, M.~Wang, D.~Guo, H.~Hao, and Q.~Liu, ``Self-mixing vibration
  measurement using emission frequency sinusoidal modulation,'' \emph{Optics
  Communications}, vol. 340, pp. 141--150, 2015.

\bibitem{Landa00}
P.~S. Landa and P.~V.~E. McClintock, ``Vibrational resonance,'' \emph{J. Phys.
  A: Math. Gen.}, vol.~33, no.~45, pp. L433--L438, 2000.

\bibitem{Chizhevsky03}
V.~N. Chizhevsky, E.~Smeu, and G.~Giacomelli, ``Experimental evidence of
  "vibrational resonance" in an optical system,'' \emph{Phys. Rev. Lett.},
  vol.~91, no.~22, p. 220602, 2003.

\bibitem{Chizhevsky12}
V.~N. Chizhevsky, ``Enhancement of response of a bistable vcsel to modulated
  orthogonal optical feedback by vibrational resonance,'' \emph{Opt. Lett.},
  vol.~37, no.~21, pp. 4386--4388, 2012.

\bibitem{Chizhevsky18}
------, ``Amplification of an autodyne signal in a bistable vertical-cavity
  surface-emitting laser with the use of a vibrational resonance,''
  \emph{Technical Physics Letters}, vol.~44, pp. 17--19, 2018.

\end{thebibliography}

\end{document}